\documentclass[review,3p,authoryear]{elsarticle}
\usepackage{titlesec}
\usepackage{verbatim}
\usepackage{amsmath}
\usepackage{setspace}
\usepackage{fancyhdr}
\usepackage[colorlinks=true]{hyperref}
\usepackage{tikz}
\usepackage{pgfplots}
\usepackage{subfigure}
\usepackage[font=footnotesize,bf]{caption}

\biboptions{square}

\fancypagestyle{plain}{%
  \fancyhf{}
  \fancyhead[L]{\sc Barnes \textit{\&} Lehman}
  \fancyhead[R]{\sc BSE in Loops}
  \cfoot{\thepage}
  
  }
\newcommand{\todo}[1]{}
\newcommand{\D}[1]{}
\newcommand{\NOTE}[1]{}

\titleformat{\section}[runin]{\normalfont\scshape\bf}{}{0pt}{}[.]
\titleformat{\subsection}[runin]{\normalfont\scshape\bf}{}{0pt}{}[.]
\titleformat{\appendix}[runin]{\normalfont\scshape\bf}{}{0pt}{}

\hypersetup{pdfauthor={Richard Barnes (ORCID: 0000-0002-0204-6040), Clarence Lehman},%
            pdftitle={Modeling of Bovine Spongiform Encephalopathy in a Two-Species Feedback Loop},%
            pdfkeywords={disease dynamics, mad cow disease, BSE, transmissible spongiform encephalopathy, prion},%
            pdfproducer={LaTeX},%
            pdfcreator={pdfLaTeX}
}

\begin{document}
\thispagestyle{empty}

{\singlespace
\noindent Cite as: Barnes and Lehman. ``Modeling of bovine spongiform encephalopathy in a two-species feedback loop". Epidemics. Vol.\ 5, Issue 2, June 2013, pp 85--91. doi: ``10.1016/j.epidem.2013.04.001".
}

\begin{frontmatter}
	\title{\sc Modeling of \\ Bovine Spongiform Encephalopathy \\ in a Two-Species Feedback Loop}
	%Searching "allintitle: mad cow disease" on Google Scholar revealed 528 hits.
	%Searching "allintitle: bovine spongiform" on Google Scholar revealed 1,840 hits.

	\author[rb]{Richard Barnes\corref{cor_rb}}
	\ead{rbarnes@umn.edu}
	\address[rb]{Ecology, Evolution, \& Behavior, University of Minnesota, USA}
	\cortext[cor_rb]{Corresponding author. ORCID: 0000-0002-0204-6040}

	\author[cl]{Clarence Lehman\corref{cor_cl}}
	\ead{lehman@umn.edu}
	\address[cl]{College of Biological Sciences, University of Minnesota, USA}
	\cortext[cor_cl]{Principal corresponding author}

	\begin{abstract} %~262 words
	\noindent\singlespacing Bovine spongiform encephalopathy, otherwise known as mad cow disease, can spread when an individual cow consumes feed containing the infected tissues of another individual, forming a one-species feedback loop. Such feedback is the primary means of transmission for BSE during epidemic conditions. Following outbreaks in the European Union and elsewhere, many governments enacted legislation designed to limit the spread of such diseases via elimination or reduction of one-species feedback loops in agricultural systems. However, two-species feedback loops---those in which infectious material from one-species is consumed by a secondary species whose tissue is then consumed by the first species---were not universally prohibited and have not been studied before. Here we present a basic ecological disease model which examines the r\^ole feedback loops may play in the spread of BSE and related diseases. Our model shows that there are critical thresholds between the infection's expansion and decrease related to the lifespan of the hosts, the growth rate of the prions, and the amount of prions circulating between hosts. The ecological disease dynamics can be intrinsically oscillatory, having outbreaks as well as refractory periods which can make it appear that the disease is under control while it is still increasing. We show that non-susceptible species that have been intentionally inserted into a feedback loop to stop the spread of disease do not, strictly by themselves, guarantee its control, though they may give that appearance by increasing the refractory period of an epidemic's oscillations. We suggest ways in which age-related dynamics and cross-species coupling should be considered in continuing evaluations aimed at maintaining a safe food supply.
	\end{abstract}

%Immediately after the abstract, provide a maximum of 5 keywords, avoiding general and plural terms and multiple concepts (avoid, for example, 'and', 'of'). Be sparing with abbreviations: only abbreviations firmly established in the field may be eligible. Please note that Keywords should NOT include words that already appear in the title of the manuscript. These keywords will be used for indexing purposes.
	\begin{keyword}
	disease dynamics \sep mad cow disease \sep BSE \sep transmissible spongiform encephalopathy \sep prion \sep disease modeling
	\end{keyword}
\end{frontmatter}

\begin{comment}
\textbf{Author Addresses}

Please contact Clarence Lehman with all publication-related correspondence
\begin{center}
\begin{tabular}{ p{5cm} p{5cm}}
  $\vcenter{\strut Clarence Lehman \par \emph{lehman@umn.edu} \par 612-625-1839 \par \, \par 123 Snyder Hall \par 1474 Gortner Avenue \par St. Paul, MN 55108}$ & 
  $\vcenter{\strut Richard Barnes \par \emph{rbarnes@umn.edu} \par 321-222-7637 \par \, \par 100 Ecology Building \par 1987 Upper Buford Circle \par St. Paul, MN 55108}$  
\end{tabular}
\end{center}
\end{comment}

\section{Introduction}
Bovine spongiform encephalopathy (BSE) is a disease in which a molecule of a specific protein misfolds into a pathogenic state; this misfolded protein then amplifies by inducing similar pathogenic misfoldings in other molecules of that protein.~\citep{prusiner1997} For short-hand in this paper, we use the term ``prion" to refer only to the misfolded form of the protein. The disease leads to neurodegeneration and death. It can be transmitted when a non-infected individual consumes prion-containing tissues from an infected individual.~\citep{cummins2001,Wilesmith1992} In the late stages of the incubation period, the brain and spinal cord are known to have especially high levels of prions; prions are also known to be present in the peripheral nervous system and ileum, but to a lesser extent.~\citep{arnold2009, masujin2007, wells1998}
%If an infected animal accumulates enough prions during its lifetime to infect, on average, more than one susceptible animal ($R_0>1$), the disease will spread in the population; on the other hand, if fewer than one susceptible animal is infected on average, the disease will die out ($R_0<1$).

Consider a hypothetical disease limited to vertebrates and transmitted when a susceptible individual consumes tissues from an infected one. The spread of the disease is normally self-limiting. Prey are consumed by predators, predators become prey, and the disease propagates along the food chain until non-susceptible invertebrate decomposers take charge.

Suppose, however, that the trophic structure is not a simple food chain, but rather a food web containing a feedback loop connecting two or more vertebrate species. For example, if a scavenger and predator species are trophically linked such that individual scavengers consume some dead predators, and living predators occasionally kill and consume the scavengers, then the dynamics are wholly different. The spread of the disease can progress cyclically around the feedback loop, limited not by the number of links in the chain but only by the size of the vertebrate populations.

Feedback loops can occur not just in nature, but also in agriculture. Livestock fed restricted diets often need food supplements, such as additional protein. Soybean meal can be used for this, but animal-derived protein is another source. Because large numbers of animals are slaughtered daily, and because not all of the slaughter is marketable to humans, a fraction remains. This fraction represents a prodigious quantity of material---up to 24 million tons or more per year in the U.S. alone~\citep{kirstein1999}---that can be rendered into a diet supplement for livestock called meat-and-bone meal, among other names. Livestock may also be fed animal byproducts such as poultry litter. For example, in 2003, the state of Florida produced one million tons of poultry litter, 350,000 of which were available for use in feed.~\citep{sapkota2007}

TSEs are known to spread among livestock such as cows~\citep{prusiner1997,nathanson1997,Wilesmith1988,wells2007} and sheep~\citep{detweiler2003} when their feed is contaminated with infected tissues. To combat such spread, many countries have enacted legislation restricting what may be fed to susceptible species by either eliminating feedback loops altogether or prohibiting one-species feedback loops. Two-species loops are not, however, universally prohibited.

Our model considers a form of feedback in which prions are amplified in one species and then fed to a secondary species in which they may or may not be decreased before being fed back to the first species. Although the disease is also thought to propagate via direct maternal transmission~\citep{donnelly1997, Wilesmith1997} and via cross-contamination of feed~\citep{abrial2005, wilesmith1996b, wilesmith1996c}, the former, if it occurs, does so only at levels insufficient to maintain an epidemic~\citep{Donnelly2002,Wilesmith2010} whereas the latter has been heavily regulated. Feedback through consumption of infected tissues is the primary means of BSE transmission during epidemic conditions~\citep{cummins2001}, so it seems worthwhile to consider this in the context of a two-species loop. This is especially so given that regulations prohibiting single-species cattle loops have created interest in making up the difference by sourcing protein from other species.~\citep{meeker}

In the two-species feedback loops we consider, infectious material passes through a secondary species. In the worst-case scenario, the secondary species becomes infected and actively contributes to the growth of the disease, but it is not necessary that infection occur. There is also the possibility that the secondary species harbours infectious material for either long or short periods without ever developing symptoms.

Although it is not known whether the disease has transmitted in this way, it represents a possible means which has not been universally prohibited nor, to the best of our knowledge, considered by prior studies.

\section{Methods of analysis}
The aim of our model is to examine population dynamics and feedback loops under the most basic conditions, applying the simplest feasible model in order to expose underlying theoretical patterns and the relative importance of parameters in one- and two-species loops. Previous models have focused on different questions such as quantifying the real world risk of human and cattle exposure during and following the UK epidemic~\citep{ferguson1999, cohen2001, nunnally2003, cohen2006}, the dynamics of the UK epidemic~\citep{ferguson1997, thornley2008}, or the spread of BSE in the cells of a single individual~\citep{nowak1998,kellershohn2001}.

Instead of tracking the number of hosts that are infected, susceptible, and resistant, as is common in ecological disease models, this model simplifies that structure by tracking the total quantity of disease agents (prions in this case) resident within each host species, treating the hosts merely as an environment in which the disease exists.

Because the conditions which could lead to an epidemic are of primary interest, the model focuses on the early growth phase of the potential epidemic, when the disease would still be spreading undetected and mitigation measures would not be in effect. Individuals are not clearly symptomatic and, therefore, are neither being culled nor dying. %Given the short artificial lifespan of most cattle, this should be a realistic scenario.

Although individual animals' susceptibility may vary, the model focuses on a subpopulation wherein all members are susceptible, and equally so. For simplicity, the model also assumes---as may be the case in an agricultural setting---that the lifespan of all hosts is artificially limited to a fixed number of years, that the size of the population is held constant, and that all individuals in an age-class are treated equally.

\subsection{One-species model}
The tightest possible feedback loop occurs when individual animals ingest tissues or byproducts of their own species, a practice which led to the spread of BSE in the United Kingdom and elsewhere.~\citep{Wilesmith1988,Wilesmith1992} We show that for such a loop, there are critical combinations of lifespan, infectivity, and feedback below which the number of infections in a population will decrease with time and eventually vanish, and above which the infection will expand epidemically.

The model uses three parameters per species, $x_i$, $c_i$, and $R$.
The net total prion level of all animals in the herd in their $i$th year of life, at the beginning of that year, is represented by $x_i$.
The prions are amplified in infected tissue by a factor of $R$ in each time unit.
Upon slaughter, some fraction $c_i$ of the tissues from the oldest age class is 
fed back and incorporated into the tissues of the younger age class $i$.
This term also incorporates the probability of infection, which relates to the infectivity of the prions, dose size, and heterogeneity of consumption.

%TODO: MERGE REVIEWERS COMMENT: I think there is a better reference for the age distribution of UK cattle then Anderson et al 1996 - the one in Donnelly, Ferguson, Ghani, Anderson proc  r soc B (2002) is better as it is more data- based and involves less inference. TODO

In the model, the lifespan of the species is $n$ years. In the United States and the United Kingdom, the lifespan of most beef cattle is held close to 2--3 years, with breeding and dairy cattle living an average of 5--7 years, but with a minority of cattle living up to 16 years.~\citep{Donnelly2002,bse_prev}
\begin{comment} From bse_prev
Age	Beef		Dairy		Total	
Age           Beef        	Beef Av		 Dairy		Dairy Av       Total		Total Av
1             6119490      6119490       4339650  4339650  10459140  10459140
2             5746000      11492000      4133000  8266000  9879000   19758000
3             3586000      10758000      2853494  8560482  6439494   19318482
4             3550589      14202356      2365992  9463968  5916581   23666324
5             3481629      17408145      1660412  8302060  5142041   25710205
6             3363550      20181300      1022177  6133062  4385727   26314362
7             3190949      22336643      583631   4085417  3774580   26422060
8             2963675      23709400      270209   2161672  3233884   25871072
9             2687158      24184422      126929   1142361  2814087   25326783
10            2372102      23721020      70138    701380   2442240   24422400
11            2033425      22367675      29732    327052   2063157   22694727
12            1688489      20261868      12542    150504   1701031   20412372
13            1354877      17613401      8731     113503   1363608   17726904
14            1048147      14674058      630      8820     1048777   14682878
15            779978       11699670      253      3795     780230    11703450
16            557086       8913376       104      1664     557190    8915040
17            381072       6478224
								6.1491118785				3.076012506			5.2161322295
\end{comment}

The model could also be arranged to include the effects of slaughtering younger age-classes and feeding them back. However, this would mean that fewer prions would make it to older age-classes leading to slower growth of the disease. The more conservative formulation of the system we examine here assumes that all members of a species make it to the same (old) age. This maximizes amplification resulting in stronger conclusions concerning the efficacy of control measures.

For cattle, the consumption of even very small amounts of infectious tissue is sufficient to spread the disease: \citet{wells2007} experimentally determined that 50\% of cattle would be clinically affected by a dose of 0.20g, with no evidence for a minimum dose. Here we arbitrarily seed our model with a ``prion level" of 1.
%Although most cattle are infected during the first two years of life \citep{anderson1996}, the disease is slow to develop.
Despite the disease's infectivity, clinical signs take an extended period to present themselves: the incubation period in cattle is approximately 3--5 years.~\citep{wells2007} \citet{nowak1998} suggest on theoretical grounds that within an individual the disease's amplification is a trade-off between the linear growth of prion aggregates and exponential growth caused by the fracturing of these aggregates, while \citet{arnold2009} experimentally determined that the disease had a doubling time of 1.2 months in the central nervous system. This implies an exponential growth rate of $R=3$ per year, which we adopt here.

Given the foregoing, if an age-class is a year long, the number of prions in each age-class in the next time unit is approximated by the following dynamical system.

{\singlespacing
\footnotesize
\begin{equation}
 \label{equ:single_species}
 \begin{pmatrix}
        x_1\cr x_2\cr x_3\cr \vdots\cr x_{n-1}\cr x_n
 \end{pmatrix}_{\displaystyle t+1}
 \kern-1em =_{\phantom{\displaystyle t+1}} \kern-1em
 \begin{pmatrix}
                0  &0  &\cdots &0  &0  &R c_1     \cr
        R  &0  &\cdots &0  &0  &R c_2     \cr
        0  &R  &\cdots &0  &0  &R c_3     \cr
        \vdots &\vdots &\ddots &\vdots &\vdots &\vdots      \cr
        0  &0  &\cdots &R  &0  &R c_{n-1} \cr
        0  &0  &\cdots &0  &R  &R c_n     \cr
 \end{pmatrix}
 \begin{pmatrix}
        x_1\cr x_2\cr x_3\cr \vdots\cr x_{n-1}\cr x_n
 \end{pmatrix}_{\displaystyle t}
\end{equation}
}

\noindent
%Or, in shorthand, $x_{t+1} = M_n x_t$.

The behavior of the system is governed by the eigenvalues of its characteristic polynomial
\begin{equation}
        \label{equ:eigen}
                 \lambda^n =   c_n     \lambda^{n-1} R
               + c_{n-1} \lambda^{n-2} R^2
               + c_{n-2} \lambda^{n-3} R^3
               + \dots
               + c_2     \lambda       R^{n-1}
               + c_1                   R^n
\end{equation}
In the discrete-time formulation of Equation~\ref{equ:single_species}, the quantity of prions, $x_i$, in each age-class will tend to increase in the population if the absolute value of any eigenvalue $\lambda$ of the matrix $M_n$ is greater than $1$. In contrast, if the absolute values of all the eigenvalues are less than $1$, the quantity of prions will tend towards zero.

Therefore, the dividing line between expansion of the disease and its extinction occurs where $\lambda=1$. Under this condition, the characteristic polynomial takes the form
%$$ n = c \sum_{i=0}^{n-1} R^i
%     = c {R^n-1\over R-1} \,. $$
%$$ \sum_{i=0}^{n-1} c_i R^{n-i} = 1 $$
\begin{equation}
        \label{equ:die_off}
%        \frac{1}{R^n} = \sum_{i=1}^{n} \frac{c_i}{R^i}
        1=R^{n+1}\sum_{i=1}^n c_i R^{-i}
\end{equation}

\begin{figure}%[h]
        \centering
	\subfigure {
	\begin{tikzpicture}[scale=0.75]
		%./mcow.py -k 10 -r 3 -n 5 -c 0.01 > /z/bob
		\begin{axis}[ylabel=Prion Level, xmin=0, xmax=20, ymin=0, width=6in, height=1.5in]
		\addplot[mark=none,domain=0:20] coordinates {
 (0.0,1.0) (0.10,1.116123) (0.20,1.245731) (0.30,1.390389) (0.40,1.551846) (0.50,1.732051) (0.60,1.933182) (0.70,2.157669) (0.80,2.408225) (0.90,2.687875) (1.0,3.0) (1.10,3.348370) (1.20,3.737193) (1.30,4.171168) (1.40,4.655537) (1.50,5.196152) (1.60,5.799546) (1.70,6.473008) (1.80,7.224674) (1.90,8.063626) (2.0,9.0) (2.10,10.045109) (2.20,11.211578) (2.30,12.513503) (2.40,13.966610) (2.50,15.588457) (2.60,17.398638) (2.70,19.419024) (2.80,21.674022) (2.90,24.190878) (3.0,27.0) (3.10,30.135326) (3.20,33.634735) (3.30,37.540508) (3.40,41.899830) (3.50,46.765372) (3.60,52.195915) (3.70,58.257071) (3.80,65.022067) (3.90,72.572635) (4.0,81.0) (4.10,90.405977) (4.20,100.904206) (4.30,112.621523) (4.40,125.699491) (4.50,140.296115) (4.60,156.587746) (4.70,174.771212) (4.80,195.06620) (4.90,217.717906) (5.0,2.430) (5.10,2.712179) (5.20,3.027126) (5.30,3.378646) (5.40,3.770985) (5.50,4.208883) (5.60,4.697632) (5.70,5.243136) (5.80,5.851986) (5.90,6.531537) (6.0,7.290) (6.10,8.136538) (6.20,9.081379) (6.30,10.135937) (6.40,11.312954) (6.50,12.626650) (6.60,14.092897) (6.70,15.729409) (6.80,17.555958) (6.90,19.594612) (7.0,21.870) (7.10,24.409614) (7.20,27.244136) (7.30,30.407811) (7.40,33.938863) (7.50,37.879951) (7.60,42.278691) (7.70,47.188227) (7.80,52.667874) (7.90,58.783835) (8.0,65.610) (8.10,73.228841) (8.20,81.732407) (8.30,91.223433) (8.40,101.816588) (8.50,113.639853) (8.60,126.836074) (8.70,141.564681) (8.80,158.003622) (8.90,176.351504) (9.0,196.830) (9.10,219.686524) (9.20,245.197221) (9.30,273.67030) (9.40,305.449764) (9.50,340.919560) (9.60,380.508222) (9.70,424.694044) (9.80,474.010865) (9.90,529.054511) (10.0,5.90490) (10.10,6.590596) (10.20,7.355917) (10.30,8.210109) (10.40,9.163493) (10.50,10.227587) (10.60,11.415247) (10.70,12.740821) (10.80,14.220326) (10.90,15.871635) (11.0,17.71470) (11.10,19.771787) (11.20,22.067750) (11.30,24.630327) (11.40,27.490479) (11.50,30.682760) (11.60,34.245740) (11.70,38.222464) (11.80,42.660978) (11.90,47.614906) (12.0,53.14410) (12.10,59.315362) (12.20,66.203250) (12.30,73.890981) (12.40,82.471436) (12.50,92.048281) (12.60,102.737220) (12.70,114.667392) (12.80,127.982933) (12.90,142.844718) (13.0,159.43230) (13.10,177.946085) (13.20,198.609749) (13.30,221.672943) (13.40,247.414309) (13.50,276.144844) (13.60,308.211660) (13.70,344.002176) (13.80,383.94880) (13.90,428.534154) (14.0,478.29690) (14.10,533.838254) (14.20,595.829247) (14.30,665.018830) (14.40,742.242927) (14.50,828.434532) (14.60,924.634979) (14.70,1032.006528) (14.80,1151.846401) (14.90,1285.602462) (15.0,14.348907) (15.10,16.015148) (15.20,17.874877) (15.30,19.950565) (15.40,22.267288) (15.50,24.853036) (15.60,27.739049) (15.70,30.960196) (15.80,34.555392) (15.90,38.568074) (16.0,43.046721) (16.10,48.045443) (16.20,53.624632) (16.30,59.851695) (16.40,66.801863) (16.50,74.559108) (16.60,83.217148) (16.70,92.880588) (16.80,103.666176) (16.90,115.704222) (17.0,129.140163) (17.10,144.136329) (17.20,160.873897) (17.30,179.555084) (17.40,200.405590) (17.50,223.677324) (17.60,249.651444) (17.70,278.641763) (17.80,310.998528) (17.90,347.112665) (18.0,387.420489) (18.10,432.408986) (18.20,482.621690) (18.30,538.665252) (18.40,601.216771) (18.50,671.031971) (18.60,748.954333) (18.70,835.925288) (18.80,932.995585) (18.90,1041.337994) (19.0,1162.261467) (19.10,1297.226958) (19.20,1447.865069) (19.30,1615.995757) (19.40,1803.650313) (19.50,2013.095913) (19.60,2246.8630) (19.70,2507.775863) (19.80,2798.986756) (19.90,3124.013982) (20.0,34.867844)
		};
		\legend{Five-year lifespan};
		\end{axis}
		\end{tikzpicture}
		}
		\subfigure{
	%./mcow.py -k 10 -r 3 -n 4 -c 0.01 > /z/bob 
	\begin{tikzpicture}[scale=0.75]
		\begin{axis}[xlabel=Time (years), ylabel=Prion Level, xmin=0, xmax=20, ymin=0, width=6in, height=1.5in]
		\addplot[mark=none,domain=0:20] coordinates {
 (0.0,1.0) (0.10,1.116123) (0.20,1.245731) (0.30,1.390389) (0.40,1.551846) (0.50,1.732051) (0.60,1.933182) (0.70,2.157669) (0.80,2.408225) (0.90,2.687875) (1.0,3.0) (1.10,3.348370) (1.20,3.737193) (1.30,4.171168) (1.40,4.655537) (1.50,5.196152) (1.60,5.799546) (1.70,6.473008) (1.80,7.224674) (1.90,8.063626) (2.0,9.0) (2.10,10.045109) (2.20,11.211578) (2.30,12.513503) (2.40,13.966610) (2.50,15.588457) (2.60,17.398638) (2.70,19.419024) (2.80,21.674022) (2.90,24.190878) (3.0,27.0) (3.10,30.135326) (3.20,33.634735) (3.30,37.540508) (3.40,41.899830) (3.50,46.765372) (3.60,52.195915) (3.70,58.257071) (3.80,65.022067) (3.90,72.572635) (4.0,0.810) (4.10,0.904060) (4.20,1.009042) (4.30,1.126215) (4.40,1.256995) (4.50,1.402961) (4.60,1.565877) (4.70,1.747712) (4.80,1.950662) (4.90,2.177179) (5.0,2.430) (5.10,2.712179) (5.20,3.027126) (5.30,3.378646) (5.40,3.770985) (5.50,4.208883) (5.60,4.697632) (5.70,5.243136) (5.80,5.851986) (5.90,6.531537) (6.0,7.290) (6.10,8.136538) (6.20,9.081379) (6.30,10.135937) (6.40,11.312954) (6.50,12.626650) (6.60,14.092897) (6.70,15.729409) (6.80,17.555958) (6.90,19.594612) (7.0,21.870) (7.10,24.409614) (7.20,27.244136) (7.30,30.407811) (7.40,33.938863) (7.50,37.879951) (7.60,42.278691) (7.70,47.188227) (7.80,52.667874) (7.90,58.783835) (8.0,0.65610) (8.10,0.732288) (8.20,0.817324) (8.30,0.912234) (8.40,1.018166) (8.50,1.136399) (8.60,1.268361) (8.70,1.415647) (8.80,1.580036) (8.90,1.763515) (9.0,1.96830) (9.10,2.196865) (9.20,2.451972) (9.30,2.736703) (9.40,3.054498) (9.50,3.409196) (9.60,3.805082) (9.70,4.246940) (9.80,4.740109) (9.90,5.290545) (10.0,5.90490) (10.10,6.590596) (10.20,7.355917) (10.30,8.210109) (10.40,9.163493) (10.50,10.227587) (10.60,11.415247) (10.70,12.740821) (10.80,14.220326) (10.90,15.871635) (11.0,17.71470) (11.10,19.771787) (11.20,22.067750) (11.30,24.630327) (11.40,27.490479) (11.50,30.682760) (11.60,34.245740) (11.70,38.222464) (11.80,42.660978) (11.90,47.614906) (12.0,0.531441) (12.10,0.593154) (12.20,0.662032) (12.30,0.738910) (12.40,0.824714) (12.50,0.920483) (12.60,1.027372) (12.70,1.146674) (12.80,1.279829) (12.90,1.428447) (13.0,1.594323) (13.10,1.779461) (13.20,1.986097) (13.30,2.216729) (13.40,2.474143) (13.50,2.761448) (13.60,3.082117) (13.70,3.440022) (13.80,3.839488) (13.90,4.285342) (14.0,4.782969) (14.10,5.338383) (14.20,5.958292) (14.30,6.650188) (14.40,7.422429) (14.50,8.284345) (14.60,9.246350) (14.70,10.320065) (14.80,11.518464) (14.90,12.856025) (15.0,14.348907) (15.10,16.015148) (15.20,17.874877) (15.30,19.950565) (15.40,22.267288) (15.50,24.853036) (15.60,27.739049) (15.70,30.960196) (15.80,34.555392) (15.90,38.568074) (16.0,0.430467) (16.10,0.480454) (16.20,0.536246) (16.30,0.598517) (16.40,0.668019) (16.50,0.745591) (16.60,0.832171) (16.70,0.928806) (16.80,1.036662) (16.90,1.157042) (17.0,1.291402) (17.10,1.441363) (17.20,1.608739) (17.30,1.795551) (17.40,2.004056) (17.50,2.236773) (17.60,2.496514) (17.70,2.786418) (17.80,3.109985) (17.90,3.471127) (18.0,3.874205) (18.10,4.324090) (18.20,4.826217) (18.30,5.386653) (18.40,6.012168) (18.50,6.710320) (18.60,7.489543) (18.70,8.359253) (18.80,9.329956) (18.90,10.413380) (19.0,11.622615) (19.10,12.972270) (19.20,14.478651) (19.30,16.159958) (19.40,18.036503) (19.50,20.130959) (19.60,22.468630) (19.70,25.077759) (19.80,27.989868) (19.90,31.240140) (20.0,0.348678)
		};
		\legend{Four-year lifespan};
		\end{axis}
	\end{tikzpicture}
		}
        \caption{Prion growth in a hypothetical animal population with a single-step feedback loop, as in Equation~\ref{equ:single_species}. Note that lifespan acts as a threshold on the spread of the disease: for the longer-lived animals (top graph) the disease spreads epidemically because the height of the peaks increases over time, but for animals living just one year less (bottom graph), the peaks decrease until the disease vanishes. Note also the presence of outbreaks and the relation of their period to the lifespan. ($R=3$, $c_1=0.01$, $c_{i\ne1}=0$, $x_{1, t=1}=1.0$, 10 time steps/year: smooth growth is assumed within each year with the prion level sampled 10 times per year to produce the figure.)}
        \label{fig:one_species}
\end{figure}
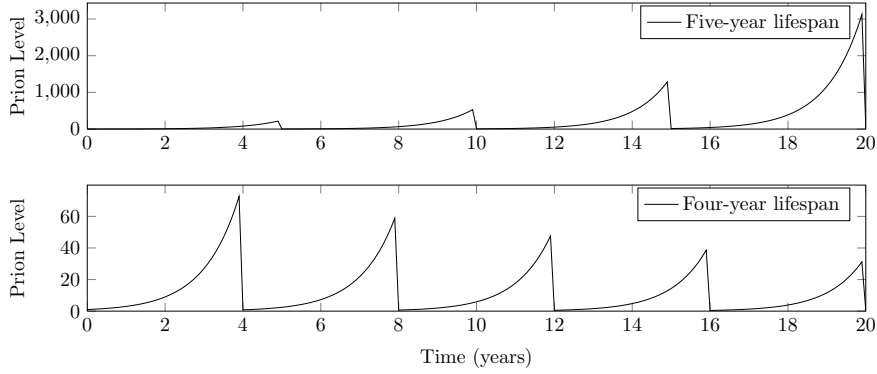
% ./a.out -r 3 -i 1 -t 100 -y 26 -l 4 -c 0.01 -c 0 -c 0 -c 0 > out_l4_t1024 
% ./a.out -r 3 -i 1 -t 100 -y 26 -l 5 -c 0.01 -c 0 -c 0 -c 0 > out_l5_t1024
 
In the simplest case, all the feedback fractions $c_i$ are zero except for one, $c_k$. This represents a situation in which animal protein supplements are given to each animal for one year only, year $k$. For instance, if $k=3$, then animals are fed supplements only in their third year of life, beginning at age 2 and ending just before age 3, as dairy cattle might be fed during their prime lactation period.\todo{Citation needed?} In such a case, Equation~\ref{equ:die_off} simplifies so that the threshold of age occurs where
\begin{equation}
        \label{equ:threshold}
        n=k-\left(1+\frac{\ln c_k}{\ln R} \right)
\end{equation}
If the lifespan, $n$, of the animals is short---less than the term to the right of the equal sign in Equation~\ref{equ:threshold}---the disease dies out. If $n$ is larger, the disease can spread. Figure~\ref{fig:one_species} shows an example of disease dynamics on both sides of the threshold, using the same initial conditions and parameters, but with differing lifespans. However, Equation~\ref{equ:die_off} is general and can be solved to find a critical threshold for any given set of parameters.
%Simulations show critical thresholds appear even with discrete daily time steps and individual-based models where age is a continuous variable and lifespan varies among individuals. %Clarence has verified that he has the code for this model.

\section{Two-species feedback loop}
%\doublespacing
If the one-step feedback loop for a long-lived Species~$L$ is broken by interposing a short-lived Species~$S$---as in the cow--pig--cow loop permitted by current U.S. regulations---the mathematics is similar to the one-species model above, but employs two embedded submatrices, one for each species, coupled by the cross-feeding between the species. For example, if one species lives for three years and the other for five, the matrix has the form %TODO: Need a citation of the U.S. regulations thing. Maybe the following?
\begin{comment}
http://www.ncbi.nlm.nih.gov/pmc/articles/PMC1867957/ :
As defined by the USDA Food Safety Inspection Service (USDA 2005b), SRMs include the skull, brain, eyes, parts of the vertebral column, spinal cord, trigeminal ganglia, and dorsal root ganglia of cattle > 30 months of age, as well as the tonsils and distal ileum of all cattle. In 1997, the FDA banned SRMs from use in cattle and other ruminant feed (GAO 2002). Nonetheless, SRMs were allowed to be incorporated into feeds for nonruminants (including poultry), and subsequent waste products from non-ruminants are still permitted in ruminant feeds (USDA 2005b).
\end{comment}

{\def\-{\hbox{-}}
\singlespacing
\begin{equation}
\footnotesize
\label{equ:two_species}
\begin{pmatrix}
        x_1\cr x_2\cr x_3\cr y_1\cr y_2\cr y_3\cr y_4\cr y_5\cr
\end{pmatrix}_{\displaystyle t+1}
 \kern-1em =_{\phantom{\displaystyle t+1}} \kern-1em
\begin{pmatrix}
                0  &0  &0    &\- &\- &\- &\- &R_L b_1  \cr
        R_S&0  &0    &\- &\- &\- &\- &R_L b_2  \cr
        0  &R_S&0    &\- &\- &\- &\- &R_L b_3  \cr
        \- &\- &R_S c_1  &0  &0  &0  &0  &0    \cr
        \- &\- &R_S c_2  &R_L&0  &0  &0  &0    \cr
        \- &\- &R_S c_3  &0  &R_L&0  &0  &0    \cr
        \- &\- &R_S c_4  &0  &0  &R_L&0  &0    \cr
        \- &\- &R_S c_5  &0  &0  &0  &R_L&0    \cr
\end{pmatrix}
\begin{pmatrix}
        x_1\cr x_2\cr x_3\cr y_1\cr y_2\cr y_3\cr y_4\cr y_5\cr
\end{pmatrix}_{\displaystyle t}
\end{equation}
}

Here, dashes (-) are the same as zeros, placed for readability. Components $x_{1,2,3}$ represent the prion levels in the three age-classes of short-lived Species $S$, while components $y_{1,2,\ldots,5}$ represent the prion levels in the five age-classes of longer-lived Species $L$. $R_S$ and $R_L$ generalize the amplification rate of the one-species case, as described above. The portion of prions fed from the oldest age-class of one species to a particular age-class $i$ of the other is encapsulated in parameters $b_i$ and $c_i$, along with the probability of infection.

\begin{figure}%[p]
    \centering
	\includegraphics[scale=1]{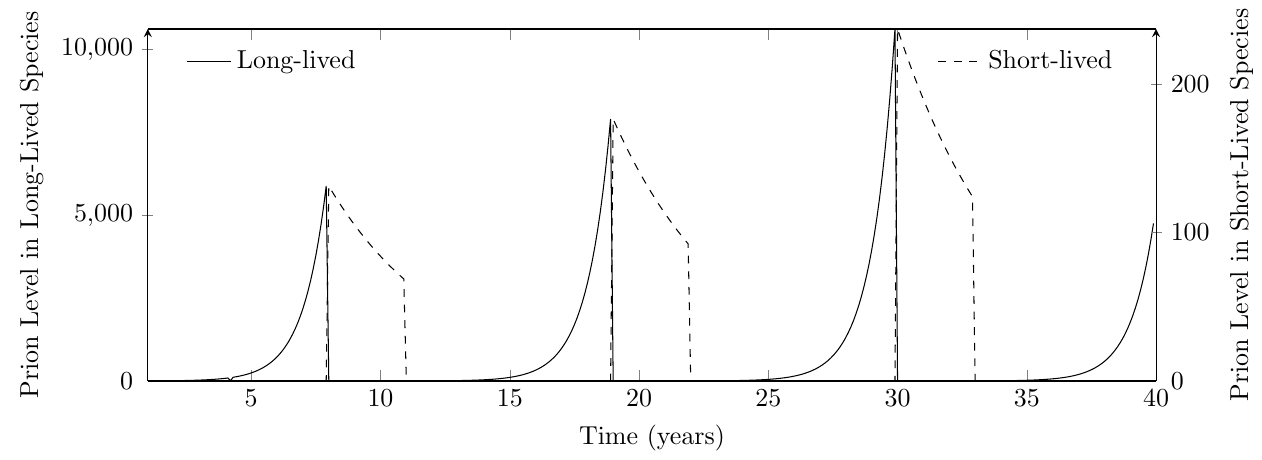}
	%./mcow2.py -k 10 -r 3 -s 0.8 -n 8 -m 3 -c 0.02,0.0,0,0,0,0,0,0,0.02 -y 40 > bob
    \caption{Prion growth in a hypothetical population with a two-species food web, calculated numerically from Equation~\ref{equ:two_species}. Growth in the susceptible species is controlled by passing prions to a non-susceptible species in which they decrease; however, under the right conditions this coupling allows the disease to spread through both, maintaining infectivity levels in the non-susceptible species. Equation~\ref{equ:two_thresh} predicts a lifespan threshold at $n\approx7.3$ years. ($R_s=0.8$, $R_l=3$, $b_{1}=c_{1}=0.02$, $b_{i\ne1}=c_{i\ne1}=0$, $y_{1,t=1}=1.0$, $m=3$, $n=8$, 10 time steps/year, where $m$ and $n$ are the life-spans of the short- and long-lived species, respectively.)}
    \label{fig:two_species}
\end{figure}

Figure~\ref{fig:two_species} depicts a numerical solution to Equation~\ref{equ:two_species} in which prions amplify in the long-lived species and decrease in the short-lived species. This demonstrates that a two-species loop does not in and of itself prevent amplification nor eliminate the possibility of spread.

%./a.out -l 3 -L 8 -r 0.3 -R 3 -c 0.02 -c 0.02 -c 0.02 -c 0.02 -c 0.02 -c 0.02 -c 0.02 -c 0.02 -c 0.02 -c 0.02 -c 0.02 -y 50 -i 1 > zz

More generally, it can be shown that the characteristic polynomial of the generalized two-species system with eigenvalue $\lambda=1$ takes the form
\begin{equation}
        \label{equ:tsg}
%        \frac{1}{{R_S}^m} = A_L \sum_{i=1}^{m} \frac{b_i}{{R_S}^i}
%        1=\sum_{j=1}^m b_j {R_S}^{m-j} \sum_{i=1}^n c_i {R_L}^{n-i}
        1={R_S}^{m+1} {R_L}^{n+1}\sum_{j=1}^m b_j {R_S}^{-j} \sum_{i=1}^n c_i {R_L}^{-i}
\end{equation}
where $m$ is the lifespan of the short-lived species, $R_S$ is the amplification factor of that species, $b_i$ represents the infectivity of the prions and the dose being fed back from a susceptible long-lived species, and the other variables are as before. From this it follows that the two-species system has a critical threshold for every combination of amplification, feedback, and lifespan.

\section{Discussion}
In response to the proven risk of the single-species loop, the European Union prohibits the incorporation of animal protein in any farmed livestock feed. The United States bans mammalian protein in ruminant feed, excluding
(a)~blood and blood products,
(b)~inspected meats used for human food and then heat processed for feed, and
(c)~meat consisting entirely of swine, horse, or poultry protein.~\citep{gao_bse}
Both regulatory structures have provisions intended to prevent cross-contamination of feed and require separation and safe disposal of certain high-risk materials known to concentrate BSE, such as brains, eyes, spinal columns, and distal ilea.\footnote{See 21CFR589.2000, 21CFR589.2001, and Regulation (EC) No 999/2001}

The United States regulatory structure does not prohibit a two-species loop in which ruminant protein is fed to pigs, horses, or poultry and their protein subsequently back to ruminants. Above, it was shown that the two-species model has critical thresholds beyond which the disease may expand in both the short- and long-lived species, with the longer-lived species establishing and maintaining the infection. The risk of this occurring in our model may be assessed by considering the sensitivity of the disease's spread to variation in its amplification rate, the amount of infectious material fed back through a loop, when this feedback occurs, and the induced lifespans of the species involved.

\begin{figure}
    \centering
	\includegraphics[scale=1]{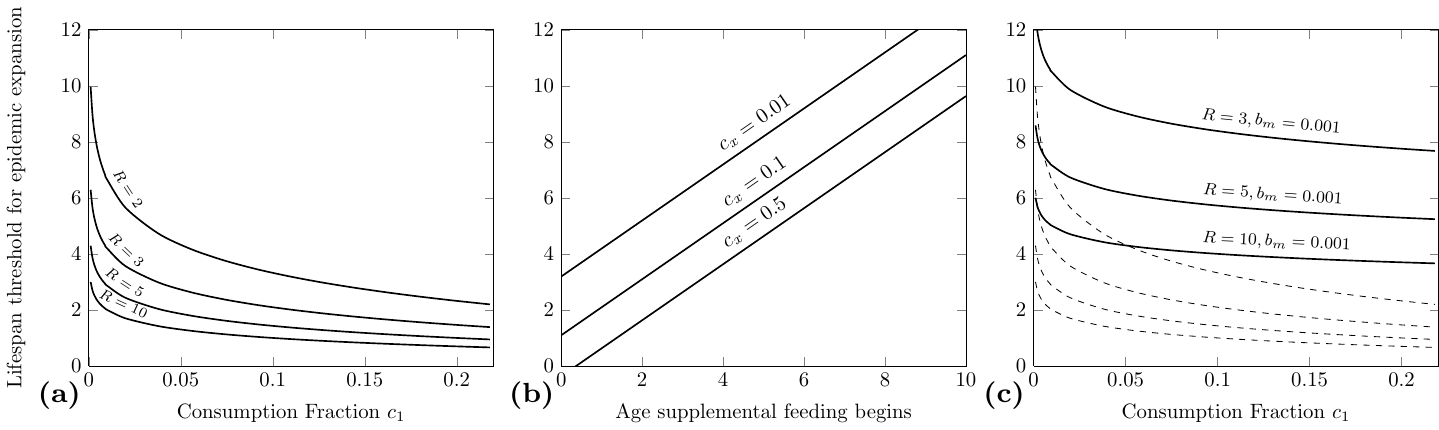}
	\caption{Sensitivity of threshold values. (a)~shows the sensitivity of the single-species loop described by Equation~\ref{equ:threshold} to variation in the feedback fraction $c_1$, assuming feedback is strictly between the oldest and youngest age-classes. (b)~shows the sensitivity of the single-species loop to variation in $k$, the initial age at which infected material is fed back, assuming that $R=3$. (c)~shows the sensitivity of the two-species loop described by Equation~\ref{equ:two_thresh} assuming a worst case wherein feed flows from the oldest age-class of the susceptible species to the oldest age-class of the nonsusceptible species and, from there, back to the youngest age-class of the susceptible species. The feedback from the susceptible to the non-susceptible species is set at $b_m=0.001$ and the amplification rate of the short-lived species is set at $R_S=1$. The sensitivity graphs of the two-species loop are overlaid on the sensitivity graphs of the single-species loop.}
	\label{fig:cthresh_one}
\end{figure}

In the one-species loop, for small values of $c$, the threshold of Equation~\ref{equ:die_off} is approximated by considering only the terms corresponding to the lowest age-class with a non-zero $c$ value; this results in Equation~\ref{equ:threshold}. Figure~\ref{fig:cthresh_one}a shows the sensitivity of the threshold to variation in the first age-class' feedback amount $c_1$ and implies that even a feedback value as low as 1\% is sufficient to drive epidemic expansion for moderate amplification values in populations maintained at 4--5 years ($R=\{5,3\}$, respectively). In contrast, for populations maintained at two years---as is the case with much of the U.S.\ herd~\citep{bse_prev}---effective separation can prevent epidemic growth for a range of amplification values in this model, provided materials from older age-classes can reliably be kept separate.

%TODO: Should I comment on this "predicting" the UK epidemic?

Figure~\ref{fig:cthresh_one}b shows the sensitivity of the threshold to variation in the age at which supplemental feeding begins, or, along the other axis, to variation in the induced lifespan. Increasing the age of initial feeding or decreasing the induced lifespan prove to be the most effective ways to reduce the growth of the disease, yielding a linear response across all values. In contrast, varying the feedback amount $c$ requires exponentially more effort to achieve increasing levels of safety.

In a two-species feedback loop, the threshold is given by solving Equation~\ref{equ:tsg}. If the feedback values are relatively small, then the system's behavior is dominated by the youngest age-class of the susceptible species and the oldest age-class of the non-susceptible species to receive infectious material. This configuration is also the minimal level of intercession for a secondary species in this model and, as such, is the worst case; other configurations will result in slower growth of the disease. In this scenario, the threshold is approximated by
\begin{comment}
\begin{equation}
\begin{alignedat}{1}
	1&={R_S}^{m+1} {R_L}^{n+1}\sum_{j=1}^m b_j {R_S}^{-j} \sum_{i=1}^n c_i {R_L}^{-i} \\
	1&={R_S}^{m+1} {R_L}^{n+1} b_m {R_S}^{-m} c_1 {R_L}^{-1} \\
	1&=R_S {R_L}^{n} b_m c_1 \\
	1&=R_S {R_L}^{n} c^2 \\
\end{alignedat}
\end{equation}

\begin{equation}
\begin{alignedat}{1}
	1&={R_S}^{m+1} {R_L}^{n+1}\sum_{j=1}^m b_j {R_S}^{-j} \sum_{i=1}^n c_i {R_L}^{-i} \\
	1&={R_S}^{m+1} {R_L}^{n+1} b_j {R_S}^{-j} c_i {R_L}^{-i} \\
	1&={R_S}^{m+1-j} {R_L}^{n+1-i} b_j c_i \\
	1&=R_S {R_L}^{n} c^2 \\
\end{alignedat}
\end{equation}
\end{comment}
\begin{equation}
\label{equ:two_thresh}
n \approx \frac{-\ln R_S-\ln(b_m c_1)}{\ln R_L}
\end{equation}
Figure~\ref{fig:cthresh_one}c shows the sensitivity of the two-species loop to variation in $c_1$, the feedback from the non-susceptible to the susceptible species. Feedback from the susceptible to the non-susceptible species is assumed to be small due to mandated separation of specified high-risk materials, and further reduced by an ``inter-species barrier" to transmission.~\citep{Wells2003} Once in the secondary species, the scenario depicted in this figure assumes that the prion level remains constant ($R_S=1$) and that the prions are introduced only in the non-susceptible species' final year of life. This limiting case favors the growth of the disease, yet even so, for low feedback values the age threshold is elevated to twice that of the single-species loop, or more. In summary, effective separation in a single-species loop is useful in reducing the possibility of growth, but does not eliminate it; thorough separation prior to feedback between two species is much more effective.

The distribution of ages in the U.S.\ herd is markedly weighted towards younger animals, with sharp drop-offs in population levels following both the first and second year in both the beef and dairy herds.~\citep{bse_prev} If all animals fed supplements draw from a common pool, this implies that the majority of infectious material ends up in short loops where the infection is not sustainable, provided there is efficient separation of infectious materials from carcasses. This would reduce the possibility of epidemic expansion in both the one- and two-species loops.

Similarly, the dose size any given animal receives is related to the homogeneity with which infectious material is mixed with non-infectious material in the feed production process. Higher degrees of homogeneity will correspond to smaller values of $c$ and $b$.

It is important to note that the dynamics of the two-species loop do not depend on the secondary species actually contracting the disease, a possibility that is still hypothetical in the cow--pig--cow loop. The sole demand is that the disease be resident long enough for the secondary species to be fed back to the first. It is possible that a secondary species could consume and passively carry infectious material for long periods without ever metabolizing it or developing visible symptoms ($R_S=1$), as can be the case with heavy metals. The infectious material may also degrade in the secondary species ($R_S<1$), or be present only present for the amount of time it takes it to pass through the secondary species' gastrointestinal tract ($R_S\approx1, b\approx0$).

Both Figure~\ref{fig:one_species} and \ref{fig:two_species} show cyclic fluctuations in prion levels over time. This is not an artifact of the model, but inherent to the nature of the disease and the feedbacks. In an SIR model, infected individuals may coexist with the susceptibles they infect, but this is not the case here: the infection of new individuals must coincide with the death of the individual that infects them. If this death is accompanied by a reduction in the net prion level and feedback is restricted to a subset of the age-classes, a cyclic pattern emerges in the early stages of the epidemic. In the case of the single-species loop, such cycles have a period equal to the induced lifespan of the species. In the case of the two-species loop, the cycles have a period equal to the combined lifespans of the two species.

\begin{figure}%[p]
    \centering
	\includegraphics[scale=1]{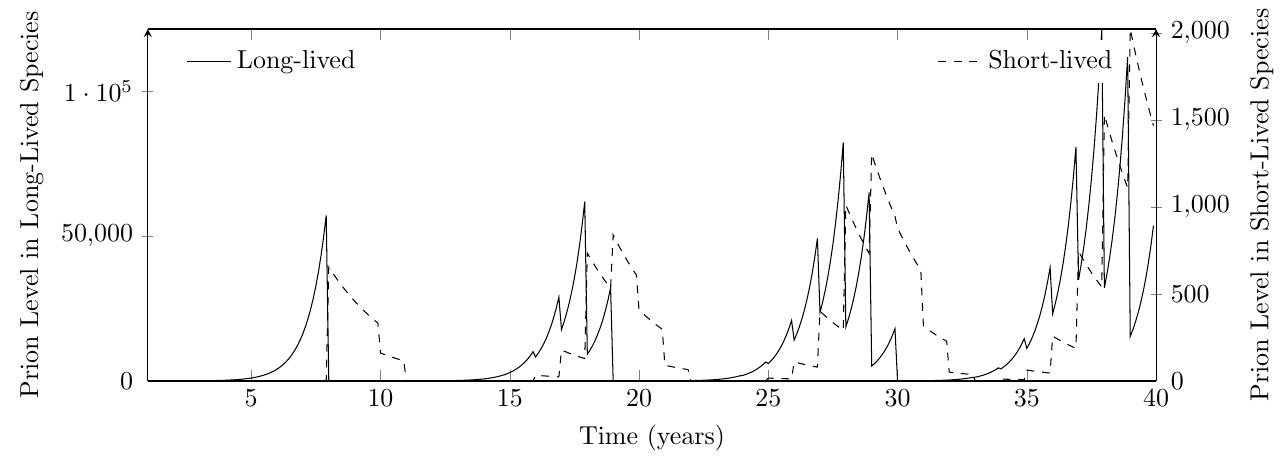}
	%./mcow2.py -k 10 -r 4 -s 0.7 -n 8 -m 3 -c 0.005,0.005,0.005,0.00,0.00,0.00,0.00,0.00,0.005,0.005,0.0 -y 40 > bob
    \caption{Prion growth in a hypothetical population with a two-species food web, calculated numerically from Equation~\ref{equ:two_species}. Although one of the species is non-susceptible, the disease spreads through both. Infectious material is fed to multiple age-classes of both species. The cyclic fluctuations and cycle period shown in Figure~\ref{fig:two_species} remain, but become increasingly dispersed and complex as the disease spreads to more and more age-classes. ($R_s=0.7$, $R_l=4$, $b_{i=1,2}=c_{i=1,2,3}=0.005$, $b_{i\ne1,2}=c_{i\ne1,2,3}=0$, $y_{1,t=1}=1.0$, $m=3$, $n=8$, 10 time steps/year.)}
    \label{fig:diffuse_cycles}
\end{figure}

In the more general scenario of contaminated materials being fed back to multiple age-classes, cycles are still present but become increasingly dispersed over time, as shown in Figure~\ref{fig:diffuse_cycles}. The more age-classes which are simultaneously exposed to the infection, the greater this dispersion is. Still, in a relatively-regimented situation such as is depicted in the figure, the behavior of the system is dominated by the youngest age-class of the susceptible species and the oldest age-class of the non-susceptible species to be fed contaminated material and the peaks of the cycles remain roughly the same initially, as Equation~\ref{equ:tsg} suggests.

A lesson to be drawn from the single-species example is that, if a species never develops a transmissible spongiform encephalopathy or related disease, it may be that
(1)~the species is not susceptible to such diseases---that is, prions cannot be amplified within its tissues and disease symptoms do not manifest themselves;
(2)~the species does not interact with prions, though the possibility of the prions being absorbed and passively carried remains;
(3)~that the species is indeed susceptible, but that the induced lifespan is below the threshold, so the disease will not spread widely and will never be manifest in detectable quantities; or,
(4)~that the amount of material being fed between age-classes is so small as to prevent the disease from growing.

The emergence of BSE among cows may have started from a rare event, but ultimately it was the dynamics of the feeding system which allowed the disease to spread. Similar diseases may exist for other species but, for the above reasons, have not yet and may never emerge. Similarly, if a two-species system never exhibits such diseases it is not necessarily because the system is protective in an absolute sense, but that it is not conducive to spread and amplification.

\section{Conclusions}
The thoughts presented here show that it is mathematically possible for cross-species feedback to spread BSE and related diseases through a susceptible population of animals, even though a species that does not exhibit susceptibility is interposed to break the feedback loop. In the process, the non-susceptible animals could actually be contaminated with the disease, albeit at lower levels. The consequences are more than an economic issue, for humans are believed to be susceptible to the disease and may contract it by ingesting infected meat. The result is called (new) variant Creutzfeldt--Jakob disease (nvCJD).~\citep{hill1997,scott1999,prusiner1997}

In the case of agricultural livestock, the maximal solution is to eliminate all use of animal byproducts in livestock feed, similar to European regulations. Short of that, the minimal safe solution is to impose lifespan limits or reduce feedback below critical thresholds, thereby driving the disease to extinction even if it occasionally gets reintroduced.

The following steps are indicated by this study as possible ways to help push the system below these thresholds:
(1)~That animals symptomatic for the disease be prohibited in animal food supplements, though this really goes without saying.
(2)~That older animals, which will have experienced the most amplification, be eliminated from animal food supplements.
(3)~That animals with longer lifespans not be fed animal food supplements, especially in younger years.
(4)~That effective separation procedures be utilized when materials are fed both within and between species.
(5)~That quantities of animal food supplements be reduced.
(6)~That artificial limits to animal life-span be imposed.
(7)~That multi-step feedback loops be eliminated where dangers exist.

Insofar as separation procedures are effective, our model shows that two-species loops reduce risk but cannot be absolutely guaranteed to eliminate it, though adverse effects in our model generally arise only for what seem to be large parameter values. However, until these systems are well-understood, prudence requires that potential feedback loops be sought out and closely scrutinized. Lessons should be drawn from all sources, including these and other models of disease dynamics, for guidance in the on-going evaluation of regulations surrounding this class of diseases.

\def\.#1 {.~}
\section{Acknowledgments}
We are grateful to
W\.illiam Hueston,    %[for discussions on infectivity etc.]
B\.en Kerr,           %[for reading the manuscript]
A\.my Kochsiek,       %[for literature searches and discussions]
J\.ustin Konen,       %[for proof-reading]
H\.olly MacCormick,   %[for reading the manuscript]
R\.ichard McGehee,    %[for review of the model and discussions]
S\.rinand Sreevatsan, %[for discussion of interspecies prion dynamics]
and
K\.endall Thomson     %[for writing the single-species characteristic polynomial]
for help and discussion in the development of this material. 
We are also grateful to the editors and to two anonymous reviewers whose valuable comments and suggestions improved the paper considerably. 
Funding has been provided in part by a fellowship to C\.larence Lehman from the University of Minnesota's Institute on the Environment.

\section{Contributions}
Both authors contributed equally. CL conceived the modeling approach for one- and two-species loops, performed the initial simulations, and wrote the initial draft. RB further developed the mathematical analysis, incorporated a literature review, and revised the manuscript. Both authors jointly contributed to the final presentation.

\singlespace
{\footnotesize
	\bibliographystyle{elsarticle-harv.bst}
	\bibliography{refs}
}
\end{document}